\documentclass[submission,copyright,creativecommons]{eptcs}
\providecommand{\event}{FROM 2023} 

\usepackage{iftex}

\ifpdf
  \usepackage{underscore}         
  \usepackage[T1]{fontenc}        
\else
  \usepackage{breakurl}           
\fi

\usepackage{coqdoc}
\usepackage{mathpartir}
\usepackage{amssymb}
\hypersetup{colorlinks=false}

\newcommand{\vswap}[3]{(\!\!(#1\ #2)\!\!) #3}
\newcommand{\swap}[3]{(#1\ #2) #3}
\newcommand{\esub}[3]{[ #2 := #3 ] #1}
\newcommand{\metasub}[3]{\{ #2 := #3 \}#1}

\title{A Formalized Extension of the Substitution Lemma in Coq}
\author{Maria J. D. Lima
  \institute{Departamento de Ciência da Computação \\
    Universidade de Brasília, Brasília, Brazil}
  \email{majuhdl@gmail.com}
  \and
  Flávio L. C. de Moura
  \institute{Departamento de Ciência da Computação \\
    Universidade de Brasília, Brasília, Brazil}
  \email{flaviomoura@unb.br}}
\def\titlerunning{A Formalized Substitution Lemma in Coq}
\def\authorrunning{M. J. D. Lima \& F. L. C. de Moura}
\begin{document}
\maketitle

\begin{abstract}
  The substitution lemma is a renowned theorem within the realm of $\lambda$-calculus theory and concerns the interactional behaviour of the metasubstitution operation. In this work, we augment the $\lambda$-calculus's grammar with an uninterpreted explicit substitution operator, which allows the use of our framework for different calculi with explicit substitutions. Our primary contribution lies in verifying that, despite these modifications, the substitution lemma continues to remain valid. This confirmation was achieved using the Coq proof assistant. Our formalization methodology employs a nominal approach, which provides a direct implementation of the $\alpha$-equivalence concept. The strategy involved in variable renaming within the proofs presents a challenge, specially on ensuring an exploration of the implications of our extension to the grammar of the $\lambda$-calculus.
\end{abstract}

\section{Introduction}

In this work, we present a formalization of the substitution lemma \cite{barendregtLambdaCalculusIts1984a} in a general framework that extends the $\lambda$-calculus with an explicit substitution operator using the Coq proof assistant \cite{teamCoqProofAssistant2021}. The source code is publicly available at

\begin{center}
\url{https://flaviomoura.info/files/msubst.v}
\end{center}

The substitution lemma is an important result concerning the composition of the substitution operation, and is usually presented as follows in the context of the $\lambda$-calculus:\\ \\

\fbox{
  \parbox{0.9\linewidth}{
 Let $t,u$ and $v$ be $\lambda$-terms, $x \neq y$ and $x\notin FV(v)$, where $FV(v)$ is the set of free variables of $v$. Then $\metasub{\metasub{t}{x}{u}}{y}{v} = \metasub{\metasub{t}{y}{v}}{x}{\metasub{u}{y}{v}}$. } }\\ \\

This is a well known result already formalized in the context of the $\lambda$-calculus \cite{berghoferHeadtoHeadComparisonBruijn2007}. Nevertheless, in the context of $\lambda$-calculi with explicit substitutions its formalization is not trivial due to the interaction between the metasubstitution and the explicit substitution operator. Our formalization is done in a nominal setting that uses the MetaLib\footnote{\url{https://github.com/plclub/metalib}} package of Coq, but no particular explicit substitution calculi is taken into account because the expected behaviour between the metasubstitution operation with the explicit substitutition constructor is the same regardless the calculus. The formalization was done with Coq (platform) version 8.15.2, which already comes with the Metalib package.  The novel contributions of this work are twofold:

\begin{enumerate}
\item The formalization is modular in the sense that no particular calculi with explicit substitutions is taken into account. Therefore, we believe that this formalization could be seen as a generic framework for proving properties of these calculi that uses the substitution lemma in the nominal setting \cite{kesnerPerpetualityFullSafe2008,nakazawaCompositionalConfluenceProofs2016,nakazawaPropertyShufflingCalculus2023};
\item  A solution to a circularity problem in the proofs is given. It adds an axiom to the formalization that allow a rewrite step inside a let expression. Such a rewrite step is problematic and does not seem to have a trivial solution.
\end{enumerate}

\section{A syntactic extension of the $\lambda$-calculus}

In this section, we present the framework of the formalization, which is based on a nominal approach \cite{gabbayNewApproachAbstract2002} where variables use names. In the nominal setting, variables are represented by atoms that are structureless entities with a decidable equality: 

\begin{verbatim}
Parameter eq_dec : forall x y : atom, {x = y} + {x <> y}.
\end{verbatim}

\noindent therefore different names mean different atoms and different variables. The nominal approach is close to the usual paper and pencil notation used in $\lambda$-calculus, whose grammar of terms is given by:

\begin{equation}\label{lambda:grammar}
 t ::= x \mid \lambda_x.t \mid t\ t
\end{equation}

\noindent where $x$ represents a variable which is taken from an enumerable set, $\lambda_x.t$ is an abstraction, and $t\ t$ is an application. The abstraction is the only binding operator: in the expression $\lambda_x.t$, $x$ binds in $t$, called the scope of the abstraction. This means that all free occurrence of $x$ in $t$ is bound in $\lambda_x.t$. A variable that is not in the scope of an abstraction is free. A variable in a term is either bound or free, but note that a varible can occur both bound and free in a term, as in $(\lambda_y. y)\ y$.

The main rule of the $\lambda$-calculus, named $\beta$-reduction, is given by:

\begin{equation}\label{lambda:beta}
 (\lambda_x.t)\ u \to_{\beta} \metasub{t}{x}{u}
\end{equation}
\noindent where $\metasub{t}{x}{u}$ represents the result of substituting all free occurrences of variable $x$ in $t$ with $u$ in such a way that renaming of bound variable may be done in order to avoid the variable capture of free variables. We call $t$ the body of the metasubstitution, and $u$ its argument. In other words, $\metasub{t}{x}{u}$ is a metanotation for a capture free substitution. For instance, the $\lambda$-term $(\lambda_x\lambda_y.x\ y)\ y$ has both bound and free occurrences of the variable $y$, and in order to $\beta$-reduce it, one has to replace (or substitute) the free variable $y$ for all free occurrences of the variable $x$ in the term $(\lambda_y.x\ y)$. But a straight substitution will capture the free variable $y$, {\it i.e.} this means that the free occurrence of $y$ before the $\beta$-reduction will become bound after the $\beta$-reduction step. A renaming of bound variables may be done to avoid such a capture, so in this example, one can take an $\alpha$-equivalent\footnote{A formal definition of this notion will be given later in this section.} term, say $(\lambda_z.x\ z)$, and perform the $\beta$-step correctly as $(\lambda_x\lambda_y.x\ y)\ y \to_{\beta} \lambda_z.y\ z$. Renaming of variables in the nominal setting is done via a name-swapping, which is formally defined as follows: \\

$\vswap{x}{y}{z} := \left\{ \begin{array}{ll}
y, & \mbox{ if } z = x; \\
x, & \mbox{ if } z = y; \\
z, & \mbox{ otherwise. } \\
\end{array}\right.$ \\

This notion can be extended to $\lambda$-terms in a straightfoward way:

\begin{equation}\label{def:swap}
\swap{x}{y}{t} := \left\{ \begin{array}{ll}
\vswap{x}{y}{z}, & \mbox{ if } t = z; \\
\lambda_{\vswap{x}{y}{z}}.\swap{x}{y}{t_1}, & \mbox{ if } t = \lambda_z.t_1; \\
\swap{x}{y}{t_1}\ \swap{x}{y}{t_2}, & \mbox{ if } t = t_1\ t_2\\
\end{array}\right.
\end{equation}

In the previous example, one could apply a swap to avoid the variable capture in a way that, a swap is applied to the body of the abstraction before applying the metasubstitution to it: $(\lambda_x\lambda_y.x\ y)\ y \to_{\beta} \metasub{(\swap{y}{z}{(\lambda_y.x\ y)})}{x}{y} = \metasub{(\lambda_z.x\ z)}{x}{y} = \lambda_z.y\ z$. Could we have used a variable substitution instead of a swapping in the previous example? Absolutely. We could have done the reduction as $(\lambda_x\lambda_y.x\ y)\ y \to_{\beta} \metasub{(\metasub{(\lambda_y.x\ y)}{y}{z})}{x}{y} = \metasub{(\lambda_z.x\ z)}{x}{y} = \lambda_z.y\ z$, but as we will shortly see, variable substitution is not stable modulo $\alpha$-equivalence, while the swapping is, thereby rendering it a more fitting choice when operating with $\alpha$-classes.

In what follows, we will adopt a mixed-notation approach, intertwining metanotation with the equivalent Coq notation. This strategy aids in elucidating the proof steps of the upcoming lemmas, enabling a clearer and more detailed comprehension of each stage in the argumentation. The corresponding Coq code for the swapping of variables, named \coqdocvar{vswap}, is defined as follows: 
\begin{coqdoccode}
\coqdocemptyline
\coqdocnoindent
\coqdockw{Definition} \coqdocvar{vswap} (\coqdocvar{x}:\coqdocvar{atom}) (\coqdocvar{y}:\coqdocvar{atom}) (\coqdocvar{z}:\coqdocvar{atom}) := \coqdockw{if} (\coqdocvar{z} == \coqdocvar{x}) \coqdockw{then} \coqdocvar{y} \coqdockw{else} \coqdockw{if} (\coqdocvar{z} == \coqdocvar{y}) \coqdockw{then} \coqdocvar{x} \coqdockw{else} \coqdocvar{z}.\coqdoceol
\coqdocemptyline
\end{coqdoccode}
\noindent therefore, the swap $\vswap{x}{y}{z}$ is written in Coq as \coqdocvar{vswap} \coqdocvar{x} \coqdocvar{y} \coqdocvar{z}. As a short example to acquaint ourselves with the Coq notation, let us show how we will write the proofs:
\begin{coqdoccode}
\coqdocemptyline
\coqdocnoindent
\coqdockw{Lemma} \coqdocvar{vswap\_id}: \coqdockw{\ensuremath{\forall}} \coqdocvar{x} \coqdocvar{y}, \coqdocvar{vswap} \coqdocvar{x} \coqdocvar{x} \coqdocvar{y} = \coqdocvar{y}.\coqdoceol
\end{coqdoccode}
\noindent{\bf Proof.} The proof is by case analysis, and it is straightforward in both cases, when \coqdocvar{x} = \coqdocvar{y} and \coqdocvar{x} \ensuremath{\not=} \coqdocvar{y}. $\hfill\Box$ 
\begin{coqdoccode}
\coqdocemptyline
\end{coqdoccode}
\subsection{An explicit substitution operator}

 The extension of the swap operation to terms require an additional comment because we will not work with the grammar (\ref{lambda:grammar}), but rather, we will extend it with an explicit substitution operator:

\begin{equation}\label{es:grammar}
  t ::= x \mid \lambda_x.t \mid t\ t \mid \esub{t}{x}{u}
\end{equation}
\noindent where $[x := u] t$ represents a term with an operator that will be evaluated with specific rules of a substitution calculus. The intended meaning of the explicit substitution is that it will simulate the metasubstitution. This formalization aims to be a generic framework applicable to any calculi with explicit substitutions using a named notation for variables. Therefore, we will not specify rules about how one can simulate the metasubstitution, but it is important to be aware that this is not a trivial task as one can easily lose important properties of the original $\lambda$-calculus \cite{melliesTypedLcalculiExplicit1995b,guillaumeCalculusDoesNot2000}.

Calculi with explicit substitutions are formalisms that deconstruct the metasubstitution operation into finer-grained steps, thereby functioning as an intermediary between the $\lambda$-calculus and its practical implementations. In other words, these calculi shed light on the execution models of higher-order languages. In fact, the development of a calculus with explicit substitutions faithful to the $\lambda$-calculus, in the sense of the preservation of some desired properties were the main motivation for such a long list of calculi with explicit substitutions invented in the last decades \cite{abadiExplicitSubstitutions1991,roseExplicitSubstitutionNames2011,benaissaLnCalculusExplicit1996,curienConfluencePropertiesWeak1996,munozConfluencePreservationStrong1996,kamareddineExtendingLcalculusExplicit1997a,blooExplicitSubstitutionEdge1999,davidLambdacalculusExplicitWeakening2001,kesnerTheoryExplicitSubstitutions2009a}.

The following inductive definition corresponds to the grammar (\ref{es:grammar}), where the explicit substitution constructor, named \coqdocvar{n\_sub}, has a special notation. Instead of writing \coqdocvar{n\_sub} \coqdocvar{t} \coqdocvar{x} \coqdocvar{u}, we will write [\coqdocvar{x} := \coqdocvar{u}] \coqdocvar{t} similarly to (\ref{es:grammar}). Accordingly, \coqdocvar{n\_sexp} denotes the set of nominal $\lambda$-expressions equipped with an explicit substitution operator, which, for simplicity, we will refer to as just ``terms''. 
\begin{coqdoccode}
\coqdocemptyline
\coqdocnoindent
\coqdockw{Inductive} \coqdocvar{n\_sexp} : \coqdockw{Set} :=\coqdoceol
\coqdocnoindent
\ensuremath{|} \coqdocvar{n\_var} (\coqdocvar{x}:\coqdocvar{atom})\coqdoceol
\coqdocnoindent
\ensuremath{|} \coqdocvar{n\_abs} (\coqdocvar{x}:\coqdocvar{atom}) (\coqdocvar{t}:\coqdocvar{n\_sexp})\coqdoceol
\coqdocnoindent
\ensuremath{|} \coqdocvar{n\_app} (\coqdocvar{t1}:\coqdocvar{n\_sexp}) (\coqdocvar{t2}:\coqdocvar{n\_sexp})\coqdoceol
\coqdocnoindent
\ensuremath{|} \coqdocvar{n\_sub} (\coqdocvar{t1}:\coqdocvar{n\_sexp}) (\coqdocvar{x}:\coqdocvar{atom}) (\coqdocvar{t2}:\coqdocvar{n\_sexp}).\coqdoceol
\coqdocemptyline
\end{coqdoccode}
The \coqdocvar{size} of a term, also written as $|t|$, and the set \coqdocvar{fv\_nom} of the free variables of a term are defined as usual: 
\begin{coqdoccode}
\coqdocemptyline
\coqdocnoindent
\coqdockw{Fixpoint} \coqdocvar{size} (\coqdocvar{t} : \coqdocvar{n\_sexp}) : \coqdocvar{nat} :=\coqdoceol
\coqdocindent{1.00em}
\coqdockw{match} \coqdocvar{t} \coqdockw{with}\coqdoceol
\coqdocindent{1.00em}
\ensuremath{|} \coqdocvar{n\_var} \coqdocvar{x} \ensuremath{\Rightarrow} 1\coqdoceol
\coqdocindent{1.00em}
\ensuremath{|} \coqdocvar{n\_abs} \coqdocvar{x} \coqdocvar{t} \ensuremath{\Rightarrow} 1 + \coqdocvar{size} \coqdocvar{t}\coqdoceol
\coqdocindent{1.00em}
\ensuremath{|} \coqdocvar{n\_app} \coqdocvar{t1} \coqdocvar{t2} \ensuremath{\Rightarrow} 1 + \coqdocvar{size} \coqdocvar{t1} + \coqdocvar{size} \coqdocvar{t2}\coqdoceol
\coqdocindent{1.00em}
\ensuremath{|} \coqdocvar{n\_sub} \coqdocvar{t1} \coqdocvar{x} \coqdocvar{t2} \ensuremath{\Rightarrow} 1 + \coqdocvar{size} \coqdocvar{t1} + \coqdocvar{size} \coqdocvar{t2}\coqdoceol
\coqdocindent{1.00em}
\coqdockw{end}.\coqdoceol
\coqdocemptyline
\coqdocnoindent
\coqdockw{Fixpoint} \coqdocvar{fv\_nom} (\coqdocvar{t} : \coqdocvar{n\_sexp}) : \coqdocvar{atoms} :=\coqdoceol
\coqdocindent{1.00em}
\coqdockw{match} \coqdocvar{t} \coqdockw{with}\coqdoceol
\coqdocindent{1.00em}
\ensuremath{|} \coqdocvar{n\_var} \coqdocvar{x} \ensuremath{\Rightarrow} \{\{\coqdocvar{x}\}\}\coqdoceol
\coqdocindent{1.00em}
\ensuremath{|} \coqdocvar{n\_abs} \coqdocvar{x} \coqdocvar{t1} \ensuremath{\Rightarrow} \coqdocvar{remove} \coqdocvar{x} (\coqdocvar{fv\_nom} \coqdocvar{t1})\coqdoceol
\coqdocindent{1.00em}
\ensuremath{|} \coqdocvar{n\_app} \coqdocvar{t1} \coqdocvar{t2} \ensuremath{\Rightarrow} \coqdocvar{fv\_nom} \coqdocvar{t1} `\coqdocvar{union}` \coqdocvar{fv\_nom} \coqdocvar{t2}\coqdoceol
\coqdocindent{1.00em}
\ensuremath{|} \coqdocvar{n\_sub} \coqdocvar{t1} \coqdocvar{x} \coqdocvar{t2} \ensuremath{\Rightarrow} (\coqdocvar{remove} \coqdocvar{x} (\coqdocvar{fv\_nom} \coqdocvar{t1})) `\coqdocvar{union}` \coqdocvar{fv\_nom} \coqdocvar{t2}\coqdoceol
\coqdocindent{1.00em}
\coqdockw{end}.\coqdoceol
\coqdocemptyline
\end{coqdoccode}
The action of a permutation on a term, written $\swap{x}{y}{t}$, is inductively defined as in (\ref{def:swap}) with the additional case for the explicit substitution operator:\vspace{.5cm}

$\swap{x}{y}{t} := \left\{ \begin{array}{ll}
\vswap{x}{y}{v}, & \mbox{ if } t \mbox{ is the variable } v; \\
\lambda_{\vswap{x}{y}{z}}. \swap{x}{y}{t_1}, & \mbox{ if } t = \lambda_z.t_1; \\
\swap{x}{y}{t_1}\ \swap{x}{y}{t_2}, & \mbox{ if } t = t_1\ t_2;\\
\esub{\swap{x}{y}{t_1}}{\vswap{x}{y}{z}}{\swap{x}{y}{t_2}}, & \mbox{ if } t = \esub{t_1}{z}{t_2}.
\end{array}\right.$ \vspace{.5cm}

The corresponding Coq definition is given by the following recursive function: 
\begin{coqdoccode}
\coqdocemptyline
\coqdocnoindent
\coqdockw{Fixpoint} \coqdocvar{swap} (\coqdocvar{x}:\coqdocvar{atom}) (\coqdocvar{y}:\coqdocvar{atom}) (\coqdocvar{t}:\coqdocvar{n\_sexp}) : \coqdocvar{n\_sexp} :=\coqdoceol
\coqdocindent{1.00em}
\coqdockw{match} \coqdocvar{t} \coqdockw{with}\coqdoceol
\coqdocindent{1.00em}
\ensuremath{|} \coqdocvar{n\_var} \coqdocvar{z}     \ensuremath{\Rightarrow} \coqdocvar{n\_var} (\coqdocvar{vswap} \coqdocvar{x} \coqdocvar{y} \coqdocvar{z})\coqdoceol
\coqdocindent{1.00em}
\ensuremath{|} \coqdocvar{n\_abs} \coqdocvar{z} \coqdocvar{t1}  \ensuremath{\Rightarrow} \coqdocvar{n\_abs} (\coqdocvar{vswap} \coqdocvar{x} \coqdocvar{y} \coqdocvar{z}) (\coqdocvar{swap} \coqdocvar{x} \coqdocvar{y} \coqdocvar{t1})\coqdoceol
\coqdocindent{1.00em}
\ensuremath{|} \coqdocvar{n\_app} \coqdocvar{t1} \coqdocvar{t2} \ensuremath{\Rightarrow} \coqdocvar{n\_app} (\coqdocvar{swap} \coqdocvar{x} \coqdocvar{y} \coqdocvar{t1}) (\coqdocvar{swap} \coqdocvar{x} \coqdocvar{y} \coqdocvar{t2})\coqdoceol
\coqdocindent{1.00em}
\ensuremath{|} \coqdocvar{n\_sub} \coqdocvar{t1} \coqdocvar{z} \coqdocvar{t2} \ensuremath{\Rightarrow} \coqdocvar{n\_sub} (\coqdocvar{swap} \coqdocvar{x} \coqdocvar{y} \coqdocvar{t1}) (\coqdocvar{vswap} \coqdocvar{x} \coqdocvar{y} \coqdocvar{z}) (\coqdocvar{swap} \coqdocvar{x} \coqdocvar{y} \coqdocvar{t2})\coqdoceol
\coqdocindent{1.00em}
\coqdockw{end}.\coqdoceol
\coqdocemptyline
\coqdocemptyline
\end{coqdoccode}
The \coqdocvar{swap} function has many interesting properties, but we will focus on the ones that are more relevant to the proofs related to the substitution lemma. Nevertheless, all lemmas can be found in the source code of the formalization\footnote{\url{https://flaviomoura.info/files/msubst.v}}. The next lemmas are simple properties that are all proved by induction on the structure of term \coqdocvar{t}: 
\begin{coqdoccode}
\coqdocemptyline
\coqdocnoindent
\coqdockw{Lemma} \coqdocvar{swap\_neq}: \coqdockw{\ensuremath{\forall}} \coqdocvar{x} \coqdocvar{y} \coqdocvar{z} \coqdocvar{w}, \coqdocvar{z} \ensuremath{\not=} \coqdocvar{w} \ensuremath{\rightarrow} \coqdocvar{vswap} \coqdocvar{x} \coqdocvar{y} \coqdocvar{z} \ensuremath{\not=} \coqdocvar{vswap} \coqdocvar{x} \coqdocvar{y} \coqdocvar{w}.\coqdoceol
\coqdocemptyline
\coqdocnoindent
\coqdockw{Lemma} \coqdocvar{swap\_size\_eq} : \coqdockw{\ensuremath{\forall}} \coqdocvar{x} \coqdocvar{y} \coqdocvar{t}, \coqdocvar{size} (\coqdocvar{swap} \coqdocvar{x} \coqdocvar{y} \coqdocvar{t}) = \coqdocvar{size} \coqdocvar{t}.\coqdoceol
\coqdocemptyline
\coqdocemptyline
\coqdocnoindent
\coqdockw{Lemma} \coqdocvar{swap\_symmetric} : \coqdockw{\ensuremath{\forall}} \coqdocvar{t} \coqdocvar{x} \coqdocvar{y}, \coqdocvar{swap} \coqdocvar{x} \coqdocvar{y} \coqdocvar{t} = \coqdocvar{swap} \coqdocvar{y} \coqdocvar{x} \coqdocvar{t}.\coqdoceol
\coqdocemptyline
\coqdocemptyline
\coqdocnoindent
\coqdockw{Lemma} \coqdocvar{swap\_involutive} : \coqdockw{\ensuremath{\forall}} \coqdocvar{t} \coqdocvar{x} \coqdocvar{y}, \coqdocvar{swap} \coqdocvar{x} \coqdocvar{y} (\coqdocvar{swap} \coqdocvar{x} \coqdocvar{y} \coqdocvar{t}) = \coqdocvar{t}.\coqdoceol
\coqdocemptyline
\coqdocnoindent
\coqdockw{Lemma} \coqdocvar{shuffle\_swap} : \coqdockw{\ensuremath{\forall}} \coqdocvar{w} \coqdocvar{y} \coqdocvar{z} \coqdocvar{t}, \coqdocvar{w} \ensuremath{\not=} \coqdocvar{z} \ensuremath{\rightarrow} \coqdocvar{y} \ensuremath{\not=} \coqdocvar{z} \ensuremath{\rightarrow} (\coqdocvar{swap} \coqdocvar{w} \coqdocvar{y} (\coqdocvar{swap} \coqdocvar{y} \coqdocvar{z} \coqdocvar{t})) = (\coqdocvar{swap} \coqdocvar{w} \coqdocvar{z} (\coqdocvar{swap} \coqdocvar{w} \coqdocvar{y} \coqdocvar{t})).\coqdoceol
\coqdocemptyline
\coqdocemptyline
\coqdocnoindent
\coqdockw{Lemma} \coqdocvar{swap\_equivariance} : \coqdockw{\ensuremath{\forall}} \coqdocvar{t} \coqdocvar{x} \coqdocvar{y} \coqdocvar{z} \coqdocvar{w}, \coqdocvar{swap} \coqdocvar{x} \coqdocvar{y} (\coqdocvar{swap} \coqdocvar{z} \coqdocvar{w} \coqdocvar{t}) = \coqdocvar{swap} (\coqdocvar{vswap} \coqdocvar{x} \coqdocvar{y} \coqdocvar{z}) (\coqdocvar{vswap} \coqdocvar{x} \coqdocvar{y} \coqdocvar{w}) (\coqdocvar{swap} \coqdocvar{x} \coqdocvar{y} \coqdocvar{t}).\coqdoceol
\coqdocemptyline
\coqdocnoindent
\coqdockw{Lemma} \coqdocvar{fv\_nom\_swap} : \coqdockw{\ensuremath{\forall}} \coqdocvar{z} \coqdocvar{y} \coqdocvar{t}, \coqdocvar{z} `\coqdocvar{notin}` \coqdocvar{fv\_nom} \coqdocvar{t} \ensuremath{\rightarrow} \coqdocvar{y} `\coqdocvar{notin}` \coqdocvar{fv\_nom} (\coqdocvar{swap} \coqdocvar{y} \coqdocvar{z} \coqdocvar{t}).\coqdoceol
\coqdocemptyline
\coqdocemptyline
\end{coqdoccode}
The standard proof strategy used so far is induction on the structure of terms. Nevertheless, the builtin induction principle automatically generated in Coq for the inductive definition \coqdocvar{n\_sexp} is not strong enough due to swappings:

\begin{verbatim}
forall P :n_sexp -> Prop,
 (forall x:atom, P(n_var x)) ->
 (forall (x:atom) (t:n_sexp), P t -> P(n_abs x t)) ->
 (forall t1:n_sexp, P t1 -> forall t2:n_sexp, P t2 -> P(n_app t1 t2)) ->
 (forall t1:n_sexp, P t1 -> forall (x:atom) (t2:n_sexp), P t2 -> P([x:=t2]t1)) ->
       forall t:n_sexp, P t
\end{verbatim}

In fact, in general, the induction hypothesis in the abstraction case (resp. explicit substitution case) refers to the body \coqdocvar{t} of the abstraction (resp. \coqdocvar{t1} of the explicit substitution), while the goal involves a swap acting on the body of the abstraction (resp. explicit substitution). In order to circunvet this problem, we defined a customized induction principle based on the size of terms: 
\begin{coqdoccode}
\coqdocemptyline
\coqdocnoindent
\coqdockw{Lemma} \coqdocvar{n\_sexp\_induction}: \coqdockw{\ensuremath{\forall}} \coqdocvar{P} : \coqdocvar{n\_sexp} \ensuremath{\rightarrow} \coqdockw{Prop}, (\coqdockw{\ensuremath{\forall}} \coqdocvar{x}, \coqdocvar{P} (\coqdocvar{n\_var} \coqdocvar{x})) \ensuremath{\rightarrow}\coqdoceol
\coqdocindent{0.50em}
(\coqdockw{\ensuremath{\forall}} \coqdocvar{t1} \coqdocvar{z}, (\coqdockw{\ensuremath{\forall}} \coqdocvar{t2} \coqdocvar{x} \coqdocvar{y}, \coqdocvar{size} \coqdocvar{t2} = \coqdocvar{size} \coqdocvar{t1} \ensuremath{\rightarrow} \coqdocvar{P} (\coqdocvar{swap} \coqdocvar{x} \coqdocvar{y} \coqdocvar{t2})) \ensuremath{\rightarrow} \coqdocvar{P} (\coqdocvar{n\_abs} \coqdocvar{z} \coqdocvar{t1})) \ensuremath{\rightarrow}\coqdoceol
\coqdocindent{0.50em}
(\coqdockw{\ensuremath{\forall}} \coqdocvar{t1} \coqdocvar{t2}, \coqdocvar{P} \coqdocvar{t1} \ensuremath{\rightarrow} \coqdocvar{P} \coqdocvar{t2} \ensuremath{\rightarrow} \coqdocvar{P} (\coqdocvar{n\_app} \coqdocvar{t1} \coqdocvar{t2})) \ensuremath{\rightarrow}\coqdoceol
\coqdocindent{0.50em}
(\coqdockw{\ensuremath{\forall}} \coqdocvar{t1} \coqdocvar{t3} \coqdocvar{z}, \coqdocvar{P} \coqdocvar{t3} \ensuremath{\rightarrow} (\coqdockw{\ensuremath{\forall}} \coqdocvar{t2} \coqdocvar{x} \coqdocvar{y}, \coqdocvar{size} \coqdocvar{t2} = \coqdocvar{size} \coqdocvar{t1} \ensuremath{\rightarrow} \coqdocvar{P} (\coqdocvar{swap} \coqdocvar{x} \coqdocvar{y} \coqdocvar{t2})) \ensuremath{\rightarrow} \coqdocvar{P} (\coqdocvar{n\_sub} \coqdocvar{t1} \coqdocvar{z} \coqdocvar{t3})) \ensuremath{\rightarrow} (\coqdockw{\ensuremath{\forall}} \coqdocvar{t}, \coqdocvar{P} \coqdocvar{t}).\coqdoceol
\coqdocemptyline
\end{coqdoccode}
\noindent which states that in order to conclude that a certain property $P$ holds for all terms, we need to prove that:
\begin{enumerate}
 \item $P$ must hold for any variable;
 \item If $P$ holds for the term $\swap{x}{y}{t_2}$, where $t_1$ and $t_2$ have the same size, then it also holds for the abstraction $\lambda_z.t_1,\forall x, y, z, t_1$ and $t_2$;
 \item If $P$ holds for the terms $t_1$ and $t_2$ the it also holds for the application $t_1\ t_2$;
 \item If $P$ holds for the term $t_3$ and for the term $\swap{x}{y}{t_2}$, where $t_1$ and $t_2$ have the same size, then it also holds for the explicit substitution $\esub{t_1}{z}{t_3},\forall x, y, z, t_1, t_2$ and $t_3$.
\end{enumerate}

The following lemma is a first example of the use of the \coqdocvar{n\_sexp\_induction} principle: 
\begin{coqdoccode}
\coqdocemptyline
\coqdocnoindent
\coqdockw{Lemma} \coqdocvar{notin\_fv\_nom\_equivariance}: \coqdockw{\ensuremath{\forall}} \coqdocvar{t} \coqdocvar{x'} \coqdocvar{x} \coqdocvar{y}, \coqdocvar{x'} `\coqdocvar{notin}` \coqdocvar{fv\_nom} \coqdocvar{t} \ensuremath{\rightarrow} \coqdocvar{vswap} \coqdocvar{x} \coqdocvar{y} \coqdocvar{x'}  `\coqdocvar{notin}` \coqdocvar{fv\_nom} (\coqdocvar{swap} \coqdocvar{x} \coqdocvar{y} \coqdocvar{t}).\coqdoceol
\end{coqdoccode}
\noindent{\bf Proof.} Note that in the paper and pencil notation, this lemma states that: \newline

If $x' \notin fv\_nom(t)$ then $\vswap{x}{y}{x'} \notin fv\_nom(\swap{x}{y}{t})$.\newline

\noindent The proof is by induction on the size of the term $t$.
\begin{enumerate} \item If $t$ is a variable, say $z$, then $x' \neq z$ by hypothesis, and we need to prove that $\vswap{x}{y}{x'} \neq \vswap{x}{y}{z}$. We conclude by lemma $swap\_neq$. \item If is an abstraction, say $t = \lambda_z. t_1$, then we have by induction hypothesis that if $x' \notin \swap{x}{y}{t_2}$ then $\vswap{x_0}{y_0}{x'} \notin \swap{x_0}{y_0}{\swap{x}{y}{t_2}}$ for any term $t_2$ with the same size as $t_1$, and any variables $x, y, x_0$ and $y_0$. At this point is important to notice that an structural induction would generate an induction hypothesis with $t_1$ only, which is not strong enough to prove the goal $\vswap{x}{y}{x'} \notin fv\_nom(\swap{x}{y}{\lambda_z. t_1})$ that has $\swap{x}{y}{t_1}$ (and not $t_1$ alone!) after the propagation of the swap. In addition, we have by hypothesis that $x' \notin fv\_nom(t_1) \backslash \{z\}$. This means that either $x' = z$ or $x' \notin fv\_nom(t_1)$, and there are two subcases: \begin{enumerate} \item If $x' = z$ then the goal is $\vswap{x}{y}{z} \notin fv\_nom(\swap{x}{y}{\lambda_z. t_1}) \Leftrightarrow$ $\vswap{x}{y}{z} \notin fv\_nom(\lambda_{\vswap{x}{y}{z}}. \swap{x}{y}{t_1}) \Leftrightarrow$\newline $\vswap{x}{y}{z} \notin fv\_nom(\swap{x}{y}{t_1})\backslash \{\vswap{x}{y}{z}\}$ we are done by lemma $notin\_remove\_3$.\footnote{This is a lemma from Metalib library and it states that {\tt forall (x y : atom) (s : atoms), x = y -> y `notin` remove x s}.} \item Otherwise, $x' \notin fv\_nom(t_1)$, and we conclude using the induction hypothesis taking $x_0=x$, $y_0=y$ and the universally quantified variables $x$ and $y$ of the internal swap as the same variable (it does not matter which one). \end{enumerate} \item The application case is straightforward from the induction hypothesis. \item The case of the explicit substitution, {\it i.e.} when $t = \esub{t_1}{z}{t_2}$, we have to prove that $\vswap{x}{y}{x'} \notin fv\_nom(\swap{x}{y}{(\esub{t_1}{z}{t_2})})$. We then propagate the swap over the explicit substitution operator and show, by the definition of $fv\_nom$, we have to prove that both $\vswap{x}{y}{x'} \notin (fv\_nom (\swap{x}{y}{t_1}))\backslash \{\vswap{x}{y}{z}\}$ and $\vswap{x}{y}{x'} \notin fv\_nom (\swap{x}{y}{t_2})$. \begin{enumerate} \item In the former case, the hypothesis $x' \notin fv\_nom(t_1)\backslash \{z\}$ generates two subcases, either $x' = z$ or $x'\notin fv\_nom(t_1)$, and we conclude with the same strategy of the abstraction case. \item The later case is straightforward by the induction hypothesis. $\hfill\Box$ \end{enumerate}\end{enumerate}

 The other direction is also true, but we skip the proof that is also by induction on the size of term \coqdocvar{t}:
\begin{coqdoccode}
\coqdocemptyline
\coqdocnoindent
\coqdockw{Lemma} \coqdocvar{notin\_fv\_nom\_remove\_swap}: \coqdockw{\ensuremath{\forall}} \coqdocvar{t} \coqdocvar{x'} \coqdocvar{x} \coqdocvar{y}, \coqdocvar{vswap} \coqdocvar{x} \coqdocvar{y} \coqdocvar{x'} `\coqdocvar{notin}` \coqdocvar{fv\_nom} (\coqdocvar{swap} \coqdocvar{x} \coqdocvar{y} \coqdocvar{t}) \ensuremath{\rightarrow} \coqdocvar{x'} `\coqdocvar{notin}` \coqdocvar{fv\_nom} \coqdocvar{t}.\coqdoceol
\coqdocemptyline
\coqdocemptyline
\end{coqdoccode}
\subsection{$\alpha$-equivalence}

 As usual in the standard presentations of the $\lambda$-calculus, we work with terms modulo $\alpha$-equivalence. This means that $\lambda$-terms are identified up to renaming of bound variables. For instance, all terms $\lambda_x.x$, $\lambda_y.y$ and $\lambda_z.z$ are seen as the same term which corresponds to the identity function. Formally, the notion of $\alpha$-equivalence is defined by the following inference rules:

\begin{mathpar}
 \inferrule*[Right={({\rm\it aeq\_var})}]{~}{x =_\alpha x} \and  \inferrule*[Right={({\rm\it aeq\_abs\_same})}]{t_1 =_\alpha t_2}{\lambda_x.t_1 =_\alpha \lambda_x.t_2} 
\end{mathpar}

\begin{mathpar}
\inferrule*[Right={({\rm\it aeq\_abs\_diff})}]{x \neq y \and x \notin fv(t_2) \and t_1 =_\alpha \swap{y}{x}{t_2}}{\lambda_x.t_1 =_\alpha \lambda_y.t_2} 
\end{mathpar}

\begin{mathpar}
 \inferrule*[Right={({\rm\it aeq\_app})}]{t_1 =_\alpha t_1' \and t_2 =_\alpha t_2'}{t_1\ t_2 =_\alpha t_1'\ t_2'} \and  \inferrule*[Right={({\rm\it aeq\_sub\_same})}]{t_1 =_\alpha t_1' \and t_2 =_\alpha t_2'}{\esub{t_1}{x}{t_2} =_\alpha \esub{t_1'}{x}{t_2'}} 
\end{mathpar}

\begin{mathpar}
\inferrule*[Right={({\rm\it aeq\_sub\_diff})}]{t_2 =_\alpha t_2' \and x \neq y \and x \notin fv(t_1') \and t_1 =_\alpha \swap{y}{x}{t_1'}}{\esub{t_1}{x}{t_2} =_\alpha \esub{t_1'}{y}{t_2'}} 
\end{mathpar}

Each of these rules correspond to a constructor in the \coqdocvar{aeq} inductive definition below:
\begin{coqdoccode}
\coqdocemptyline
\coqdocnoindent
\coqdockw{Inductive} \coqdocvar{aeq} : \coqdocvar{n\_sexp} \ensuremath{\rightarrow} \coqdocvar{n\_sexp} \ensuremath{\rightarrow} \coqdockw{Prop} :=\coqdoceol
\coqdocnoindent
\ensuremath{|} \coqdocvar{aeq\_var} : \coqdockw{\ensuremath{\forall}} \coqdocvar{x}, \coqdocvar{aeq} (\coqdocvar{n\_var} \coqdocvar{x}) (\coqdocvar{n\_var} \coqdocvar{x})\coqdoceol
\coqdocnoindent
\ensuremath{|} \coqdocvar{aeq\_abs\_same} : \coqdockw{\ensuremath{\forall}} \coqdocvar{x} \coqdocvar{t1} \coqdocvar{t2}, \coqdocvar{aeq} \coqdocvar{t1} \coqdocvar{t2} \ensuremath{\rightarrow} \coqdocvar{aeq} (\coqdocvar{n\_abs} \coqdocvar{x} \coqdocvar{t1})(\coqdocvar{n\_abs} \coqdocvar{x} \coqdocvar{t2})\coqdoceol
\coqdocnoindent
\ensuremath{|} \coqdocvar{aeq\_abs\_diff} : \coqdockw{\ensuremath{\forall}} \coqdocvar{x} \coqdocvar{y} \coqdocvar{t1} \coqdocvar{t2}, \coqdocvar{x} \ensuremath{\not=} \coqdocvar{y} \ensuremath{\rightarrow} \coqdocvar{x} `\coqdocvar{notin}` \coqdocvar{fv\_nom} \coqdocvar{t2} \ensuremath{\rightarrow} \coqdocvar{aeq} \coqdocvar{t1} (\coqdocvar{swap} \coqdocvar{y} \coqdocvar{x} \coqdocvar{t2}) \ensuremath{\rightarrow}\coqdoceol
\coqdocindent{1.00em}
\coqdocvar{aeq} (\coqdocvar{n\_abs} \coqdocvar{x} \coqdocvar{t1}) (\coqdocvar{n\_abs} \coqdocvar{y} \coqdocvar{t2})\coqdoceol
\coqdocnoindent
\ensuremath{|} \coqdocvar{aeq\_app} : \coqdockw{\ensuremath{\forall}} \coqdocvar{t1} \coqdocvar{t2} \coqdocvar{t1'} \coqdocvar{t2'}, \coqdocvar{aeq} \coqdocvar{t1} \coqdocvar{t1'} \ensuremath{\rightarrow} \coqdocvar{aeq} \coqdocvar{t2} \coqdocvar{t2'} \ensuremath{\rightarrow} \coqdocvar{aeq} (\coqdocvar{n\_app} \coqdocvar{t1} \coqdocvar{t2}) (\coqdocvar{n\_app} \coqdocvar{t1'} \coqdocvar{t2'})\coqdoceol
\coqdocnoindent
\ensuremath{|} \coqdocvar{aeq\_sub\_same} : \coqdockw{\ensuremath{\forall}} \coqdocvar{t1} \coqdocvar{t2} \coqdocvar{t1'} \coqdocvar{t2'} \coqdocvar{x}, \coqdocvar{aeq} \coqdocvar{t1} \coqdocvar{t1'} \ensuremath{\rightarrow} \coqdocvar{aeq} \coqdocvar{t2} \coqdocvar{t2'} \ensuremath{\rightarrow} \coqdocvar{aeq} ([\coqdocvar{x} := \coqdocvar{t2}] \coqdocvar{t1}) ([\coqdocvar{x} := \coqdocvar{t2'}] \coqdocvar{t1'})\coqdoceol
\coqdocnoindent
\ensuremath{|} \coqdocvar{aeq\_sub\_diff} : \coqdockw{\ensuremath{\forall}} \coqdocvar{t1} \coqdocvar{t2} \coqdocvar{t1'} \coqdocvar{t2'} \coqdocvar{x} \coqdocvar{y}, \coqdocvar{aeq} \coqdocvar{t2} \coqdocvar{t2'} \ensuremath{\rightarrow} \coqdocvar{x} \ensuremath{\not=} \coqdocvar{y} \ensuremath{\rightarrow} \coqdocvar{x} `\coqdocvar{notin}` \coqdocvar{fv\_nom} \coqdocvar{t1'} \ensuremath{\rightarrow} \coqdocvar{aeq} \coqdocvar{t1} (\coqdocvar{swap} \coqdocvar{y} \coqdocvar{x} \coqdocvar{t1'}) \ensuremath{\rightarrow}\coqdoceol
\coqdocindent{1.00em}
\coqdocvar{aeq} ([\coqdocvar{x} := \coqdocvar{t2}] \coqdocvar{t1}) ([\coqdocvar{y} := \coqdocvar{t2'}] \coqdocvar{t1'}).\coqdoceol
\coqdocemptyline
\end{coqdoccode}
In what follows, we use a infix notation for $\alpha$-equivalence in the Coq code. Therefore, we write \coqdocvar{t} =\coqdocvar{a} \coqdocvar{u} instead of \coqdocvar{aeq} \coqdocvar{t} \coqdocvar{u}. The above notion defines an equivalence relation over the set \coqdocvar{n\_sexp} of nominal expressions with explicit substitutions, {\it i.e.} the \coqdocvar{aeq} relation is reflexive, symmetric and transitive (proofs in the source file\footnote{\url{https://flaviomoura.info/files/msubst.v}}). In addition, $\alpha$-equivalent terms have the same size, and the same set of free variables: 
\begin{coqdoccode}
\coqdocemptyline
\coqdocnoindent
\coqdockw{Lemma} \coqdocvar{aeq\_size}: \coqdockw{\ensuremath{\forall}} \coqdocvar{t1} \coqdocvar{t2}, \coqdocvar{t1} =\coqdocvar{a} \coqdocvar{t2} \ensuremath{\rightarrow} \coqdocvar{size} \coqdocvar{t1} = \coqdocvar{size} \coqdocvar{t2}.\coqdoceol
\coqdocemptyline
\coqdocnoindent
\coqdockw{Lemma} \coqdocvar{aeq\_fv\_nom} : \coqdockw{\ensuremath{\forall}} \coqdocvar{t1} \coqdocvar{t2}, \coqdocvar{t1} =\coqdocvar{a} \coqdocvar{t2} \ensuremath{\rightarrow} \coqdocvar{fv\_nom} \coqdocvar{t1} [=] \coqdocvar{fv\_nom} \coqdocvar{t2}.\coqdoceol
\coqdocemptyline
\coqdocemptyline
\end{coqdoccode}
The key point of the nominal approach is that the swap operation is stable under $\alpha$-equivalence in the sense that, $t_1 =_\alpha t_2$ if, and only if $\swap{x}{y}{t_1} =_\alpha \swap{x}{y}{t_2}, \forall t_1, t_2, x, y$. Note that this is not true for renaming substitutions: in fact, $\lambda_x.z =_\alpha \lambda_y.z$, but $\metasub{(\lambda_x.z)}{z}{x} = \lambda_x.x \neq_\alpha \metasub{\lambda_y.x (\lambda_y.z)}{z}{x}$, assuming that $x \neq y$. This stability result is formalized as follows:
\begin{coqdoccode}
\coqdocemptyline
\coqdocnoindent
\coqdockw{Corollary} \coqdocvar{aeq\_swap}: \coqdockw{\ensuremath{\forall}} \coqdocvar{t1} \coqdocvar{t2} \coqdocvar{x} \coqdocvar{y}, \coqdocvar{t1} =\coqdocvar{a} \coqdocvar{t2} \ensuremath{\leftrightarrow} (\coqdocvar{swap} \coqdocvar{x} \coqdocvar{y} \coqdocvar{t1}) =\coqdocvar{a} (\coqdocvar{swap} \coqdocvar{x} \coqdocvar{y} \coqdocvar{t2}).\coqdoceol
\coqdocemptyline
\coqdocemptyline
\end{coqdoccode}
When both variables in a swap do not occur free in a term, it eventually renames only bound variables, {\it i.e.} the action of this swap results in a term that is $\alpha$-equivalent to the original term. This is the content of the following lemma:
\begin{coqdoccode}
\coqdocemptyline
\coqdocnoindent
\coqdockw{Lemma} \coqdocvar{swap\_reduction}: \coqdockw{\ensuremath{\forall}} \coqdocvar{t} \coqdocvar{x} \coqdocvar{y}, \coqdocvar{x} `\coqdocvar{notin}` \coqdocvar{fv\_nom} \coqdocvar{t} \ensuremath{\rightarrow} \coqdocvar{y} `\coqdocvar{notin}` \coqdocvar{fv\_nom} \coqdocvar{t} \ensuremath{\rightarrow} (\coqdocvar{swap} \coqdocvar{x} \coqdocvar{y} \coqdocvar{t}) =\coqdocvar{a}  \coqdocvar{t}.\coqdoceol
 \coqdocemptyline
\coqdocemptyline
\end{coqdoccode}
There are several other interesting auxiliary properties that need to be proved before achieving the substitution lemma. In what follows, we refer only to the tricky or challenging ones, but the interested reader can have a detailed look in the source file. Note that, swaps are introduced in proofs by the rules $\mbox{\it aeq}\_\mbox{\it abs}\_\mbox{\it diff}$ and $\mbox{\it aeq}\_\mbox{\it sub}\_\mbox{\it diff}$. As we will see, the proof steps involving these rules are trick because a naïve strategy can easily get blocked in a branch without proof. We conclude this section, with a lemma that gives the conditions for two swaps with a common variable to be merged: 
\begin{coqdoccode}
\coqdocemptyline
\coqdocnoindent
\coqdockw{Lemma} \coqdocvar{aeq\_swap\_swap}: \coqdockw{\ensuremath{\forall}} \coqdocvar{t} \coqdocvar{x} \coqdocvar{y} \coqdocvar{z}, \coqdocvar{z} `\coqdocvar{notin}` \coqdocvar{fv\_nom} \coqdocvar{t} \ensuremath{\rightarrow} \coqdocvar{x} `\coqdocvar{notin}` \coqdocvar{fv\_nom} \coqdocvar{t} \ensuremath{\rightarrow} (\coqdocvar{swap} \coqdocvar{z} \coqdocvar{x} (\coqdocvar{swap} \coqdocvar{x} \coqdocvar{y} \coqdocvar{t})) =\coqdocvar{a} (\coqdocvar{swap} \coqdocvar{z} \coqdocvar{y} \coqdocvar{t}).\coqdoceol
\end{coqdoccode}
\noindent{\bf Proof.} Before commenting this proof, we state the lemma with the pencil and paper (meta)notation: \newline

If $z\notin fv\_nom(t)$ and $x \notin fv\_nom(t)$ then $\swap{z}{x}{\swap{x}{y}{t}} =_{\alpha} \swap{z}{y}{t}$.\newline

\noindent Initially, observe the similarity of the left hand side (LHS) of the $\alpha$-equation with the lemma \coqdocvar{shuffle\_swap}:

$\forall w\ y\ z\ t, w \neq z \to y \neq z \to \swap{w}{y}{(\swap{y}{z}{t})} = \swap{w}{z}{(\swap{w}{y}{t})}$

In order to use it, we need to have that both \coqdocvar{z} \ensuremath{\not=} \coqdocvar{y} and \coqdocvar{x} \ensuremath{\not=} \coqdocvar{y}. We start comparing \coqdocvar{z} and \coqdocvar{y}: \begin{enumerate} \item If $z = y$ then the right hand side (RHS) reduces to $t$ because the swap is trivial, and the LHS also reduces to $t$ since swap is involutive. \item When $z \neq y$ then we proceed by comparing $x$ and $y$: \begin{enumerate} \item If $x = y$ then both sides of the $\alpha$-equation reduces to $\swap{z}{y}{t}$, and we are done. \item Finally, when $x \neq y$, we can apply the lemma $shuffle\_swap$, and use lemma $aeq\_swap$ to reduce the current goal to $\swap{z}{x}{t} =_{\alpha} t$, and we conclude by lemma $swap\_reduction$ since both $z$ and $x$ are not in the set of free variables of the term $t$. $\hfill\Box$ \end{enumerate}\end{enumerate}
\begin{coqdoccode}
\coqdocemptyline
\end{coqdoccode}
\section{The metasubstitution operation of the $\lambda$-calculus}

 As presented in Section 2, the main operation of the $\lambda$-calculus is the $\beta$-reduction (\ref{lambda:beta}) that expresses how to evaluate a function applied to an argument. The $\beta$-contractum $\metasub{t}{x}{u}$ represents a capture free in the sense that no free variable becomes bound by the application of the metasubstitution. This operation is in the meta level because it is outside the grammar of the $\lambda$-calculus (and hence its name). In \cite{barendregtLambdaCalculusIts1984a}, Barendregt defines it as follows:

\vspace{.5cm}
$\metasub{t}{x}{u} = \left\{
 \begin{array}{ll}
  u, & \mbox{ if } t = x; \\
  y, & \mbox{ if } t = y \mbox{ and } x \neq y; \\
  \metasub{t_1}{x}{u}\ \metasub{t_2}{x}{u}, & \mbox{ if } t = t_1\ t_2; \\
  \lambda_y.(\metasub{t_1}{x}{u}), & \mbox{ if } t = \lambda_y.t_1.
 \end{array}\right.$ \vspace{.5cm}

\noindent where it is assumed the so called ``Barendregt's variable convention'':\\

\fbox{
  \parbox{0.9\linewidth}{
 If $t_1, t_2, \ldots, t_n$ occur in a certain mathematical context (e.g. definition, proof), then in these terms all bound variables are chosen to be different from the free variables.} }\\

This means that we are assumming that both $x \neq y$ and $y\notin fv(u)$ in the case $t = \lambda_y.t_1$. This approach is very convenient in informal proofs because it avoids having to rename bound variables. In order to formalize the capture free substitution, {\it i.e.} the metasubstitution, there are different possible approaches. In our case, we perform a renaming of bound variables whenever the metasubstitution is propagated inside a binder. In our case, there are two binders: abstractions and explicit substitutions.

Let $t$ and $u$ be terms, and $x$ a variable. The result of substituting $u$ for the free ocurrences of $x$ in $t$, written $\metasub{t}{x}{u}$ is defined as follows:\newline
\begin{equation}\label{msubst}
\metasub{t}{x}{u} = \left\{
 \begin{array}{l@{}l}
  u, & \mbox{ if } t = x; \\
  y, & \mbox{ if } t = y\ (x \neq y); \\
  \metasub{t_1}{x}{u}\ \metasub{t_2}{x}{u}, & \mbox{ if } t = t_1\ t_2; \\
  \lambda_x.t_1, & \mbox{ if } t = \lambda_x.t_1; \\
  \lambda_z.(\metasub{(\swap{y}{z}{t_1})}{x}{u}), & \mbox{ if } t = \lambda_y.t_1, x \neq y, z\notin fv(t)\cup fv(u) \cup \{x\}; \\
  \esub{t_1}{x}{\metasub{t_2}{x}{u}}, & \mbox{ if } t = \esub{t_1}{x}{t_2}; \\
  \esub{\metasub{(\swap{y}{z}{t_1})}{x}{u}}{z}{\metasub{t_2}{x}{u}}, & \mbox{ if } t = \esub{t_1}{y}{t_2}, x \neq y, z\notin fv(t)\cup fv(u) \cup \{x\}.
 \end{array}\right.
\end{equation}

\noindent and the corresponding Coq code is as follows: 
\begin{coqdoccode}
\coqdocemptyline
\coqdocnoindent
\coqdockw{Function} \coqdocvar{subst\_rec\_fun} (\coqdocvar{t}:\coqdocvar{n\_sexp}) (\coqdocvar{u} :\coqdocvar{n\_sexp}) (\coqdocvar{x}:\coqdocvar{atom}) \{\coqdockw{measure} \coqdocvar{size} \coqdocvar{t}\} : \coqdocvar{n\_sexp} :=\coqdoceol
\coqdocindent{1.00em}
\coqdockw{match} \coqdocvar{t} \coqdockw{with}\coqdoceol
\coqdocindent{1.00em}
\ensuremath{|} \coqdocvar{n\_var} \coqdocvar{y} \ensuremath{\Rightarrow} \coqdockw{if} (\coqdocvar{x} == \coqdocvar{y}) \coqdockw{then} \coqdocvar{u} \coqdockw{else} \coqdocvar{t}\coqdoceol
\coqdocindent{1.00em}
\ensuremath{|} \coqdocvar{n\_abs} \coqdocvar{y} \coqdocvar{t1} \ensuremath{\Rightarrow} \coqdockw{if} (\coqdocvar{x} == \coqdocvar{y}) \coqdockw{then} \coqdocvar{t} \coqdockw{else} \coqdockw{let} (\coqdocvar{z},\coqdocvar{\_}) :=\coqdoceol
\coqdocindent{2.00em}
\coqdocvar{atom\_fresh} (\coqdocvar{fv\_nom} \coqdocvar{u} `\coqdocvar{union}` \coqdocvar{fv\_nom} \coqdocvar{t} `\coqdocvar{union}` \{\{\coqdocvar{x}\}\}) \coqdoctac{in} \coqdocvar{n\_abs} \coqdocvar{z} (\coqdocvar{subst\_rec\_fun} (\coqdocvar{swap} \coqdocvar{y} \coqdocvar{z} \coqdocvar{t1}) \coqdocvar{u} \coqdocvar{x})\coqdoceol
\coqdocindent{1.00em}
\ensuremath{|} \coqdocvar{n\_app} \coqdocvar{t1} \coqdocvar{t2} \ensuremath{\Rightarrow} \coqdocvar{n\_app} (\coqdocvar{subst\_rec\_fun} \coqdocvar{t1} \coqdocvar{u} \coqdocvar{x}) (\coqdocvar{subst\_rec\_fun} \coqdocvar{t2} \coqdocvar{u} \coqdocvar{x})\coqdoceol
\coqdocindent{1.00em}
\ensuremath{|} \coqdocvar{n\_sub} \coqdocvar{t1} \coqdocvar{y} \coqdocvar{t2} \ensuremath{\Rightarrow} \coqdockw{if} (\coqdocvar{x} == \coqdocvar{y}) \coqdockw{then} \coqdocvar{n\_sub} \coqdocvar{t1} \coqdocvar{y} (\coqdocvar{subst\_rec\_fun} \coqdocvar{t2} \coqdocvar{u} \coqdocvar{x}) \coqdockw{else} \coqdockw{let} (\coqdocvar{z},\coqdocvar{\_}) :=\coqdoceol
\coqdocindent{2.00em}
\coqdocvar{atom\_fresh} (\coqdocvar{fv\_nom} \coqdocvar{u} `\coqdocvar{union}` \coqdocvar{fv\_nom} \coqdocvar{t} `\coqdocvar{union}` \{\{\coqdocvar{x}\}\}) \coqdoctac{in}\coqdoceol
\coqdocindent{2.00em}
\coqdocvar{n\_sub} (\coqdocvar{subst\_rec\_fun} (\coqdocvar{swap} \coqdocvar{y} \coqdocvar{z} \coqdocvar{t1}) \coqdocvar{u} \coqdocvar{x}) \coqdocvar{z} (\coqdocvar{subst\_rec\_fun} \coqdocvar{t2} \coqdocvar{u} \coqdocvar{x}) \coqdockw{end}.\coqdoceol
\coqdocemptyline
\end{coqdoccode}\coqdocemptyline\coqdocemptyline

Note that this function is not structurally recursive due to the swaps in the recursive calls, and that's why we need to provide the size of the term $t$ as the measure parameter. Alternatively, a structurally recursive version of the function \coqdocvar{subst\_rec\_fun} can be found in the file \coqdocvar{nominal.v} of the \coqdocvar{Metalib} library\footnote{\url{https://github.com/plclub/metalib}}. It has the size of the term as an explicit parameter in which the substitution will be performed, and hence one has to deal with the size of the term in each recursive call. We write \{\coqdocvar{x}:=\coqdocvar{u}\}\coqdocvar{t} instead of \coqdocvar{subst\_rec\_fun} \coqdocvar{t} \coqdocvar{u} \coqdocvar{x}, and refer to it just as ``metasubstitution''.
\begin{coqdoccode}
\coqdocemptyline
\coqdocemptyline
\end{coqdoccode}
The following lemma states that if $x \notin fv(t)$ then $\metasub{t}{x}{u} =_\alpha t$. In informal proofs the conclusion of this lemma is usually stated as a syntactic equality, {\i.e.} $\metasub{t}{x}{u} = t$ instead of the $\alpha$-equivalence, but the function \coqdocvar{subst\_rec\_fun} renames bound variables whenever the metasubstitution is propagated inside an abstraction or an explicit substitution, even in the case that the metasubstitution has no effect in the subterm it is propagated, as long as the variables of the metasubstitution and the binder (abstraction or explicit substitution) are different of each other. That's why the syntactic equality does not hold here. 
\begin{coqdoccode}
\coqdocemptyline
\coqdocnoindent
\coqdockw{Lemma} \coqdocvar{m\_subst\_notin}: \coqdockw{\ensuremath{\forall}} \coqdocvar{t} \coqdocvar{u} \coqdocvar{x}, \coqdocvar{x} `\coqdocvar{notin}` \coqdocvar{fv\_nom} \coqdocvar{t} \ensuremath{\rightarrow} \{\coqdocvar{x} := \coqdocvar{u}\}\coqdocvar{t} =\coqdocvar{a} \coqdocvar{t}.\coqdoceol
\end{coqdoccode}
\noindent{\bf Proof.} The proof is done by induction on the size of the term \coqdocvar{t} using \coqdocvar{n\_sexp\_induction} defined above. The interesting cases are the abstraction and the explicit substituion. We focus in the abstraction case, {\it i.e.} when $t = \lambda_y.t_1$, where the goal to be proven is $\metasub{(\lambda_y.t_1)}{x}{u} =_\alpha \lambda_y.t_1$. We consider two cases: \begin{enumerate} \item If $x = y$ the result is trivial because both LHS and RHS are equal to $\lambda_y.t_1$ \item If $x \neq y$ , we have to prove that $\lambda_z. \metasub{\swap{y}{z}{t_1}}{x}{u} =_{\alpha} \lambda_y. t_1$, where $z$ is a fresh name not in the set $fv\_nom(u)\cup fv\_nom(\lambda_y.t_1)\cup \{x\}$. The induction hypothesis express the fact that every term with the same size as the body $t_1$ of the abstraction  satisfies the property to be proven: $\forall t', |t'| = |t_1| \to \forall u\ x'\ x_0\ y_0, x' \notin fv(\swap{x_0}{y_0}{t'}) \to \metasub{(\swap{x_0}{y_0}{t'})}{x'}{u} =_\alpha \swap{x}{y}{t'}$. Therefore, according to the definition of the metasubstitution (function [subst_rec_fun]), the variable $y$ will be renamed to $z$, and the metasubstitution is propagated inside the abstraction resulting in the following goal: $\lambda_z.\metasub{(\swap{z}{y}{t_1})}{x}{u} =_\alpha \lambda_y.t_1$. Since $z \notin fv\_nom(\lambda_y.t_1) = fv\_nom(t_1)\backslash \{y\}$, there are two cases to consider, either $z = y$ or $z \in fv(t_1)$:
\begin{enumerate}
 \item $z = y$: In this case, we are done by the induction hypothesis taking $x_0=y_0=y$, for instance.
 \item $z \neq y$: In this case, we can apply the rule $\mbox{\it aeq}\_\mbox{\it abs}\_\mbox{\it diff}$, resulting in the goal $\metasub{(\swap{y}{z}{t_1})}{x}{u} =_\alpha \swap{y}{z}{t_1}$ which holds by the induction hypothesis, since $|\swap{z}{y}{t_1}| = |t_1|$ and $x \notin fv\_nom(\swap{y}{z}{t_1})$ because $x \neq z$, $x \neq y$ and $x \notin fv\_nom(t_1)$.
\end{enumerate}
\end{enumerate}

The explicit substitution case is also interesting, {\it i.e.} if $t = \esub{t_1}{y}{t_2}$, but it follows a similar strategy used in the abstraction case for $t_1$. For $t_2$ the result follows from the induction hypothesis. $\hfill\Box$
\begin{coqdoccode}
\coqdocemptyline
\coqdocemptyline
\end{coqdoccode}
The following lemmas concern the expected behaviour of the metasubstitution when the metasubstitution's variable is equal to the abstraction's variable. Their proofs are straightforward from the definition \coqdocvar{subst\_rec\_fun}. The corresponding version when the metasubstitution's variable is different from the abstraction's variable will be presented later. \newline
\begin{coqdoccode}
\coqdocemptyline
\coqdocnoindent
\coqdockw{Lemma} \coqdocvar{m\_subst\_abs\_eq}: \coqdockw{\ensuremath{\forall}} \coqdocvar{u} \coqdocvar{x} \coqdocvar{t}, \{\coqdocvar{x} := \coqdocvar{u}\}(\coqdocvar{n\_abs} \coqdocvar{x} \coqdocvar{t}) = \coqdocvar{n\_abs} \coqdocvar{x} \coqdocvar{t}.\coqdoceol
\coqdocemptyline\coqdocemptyline
\coqdocnoindent
\coqdockw{Lemma} \coqdocvar{m\_subst\_sub\_eq}: \coqdockw{\ensuremath{\forall}} \coqdocvar{u} \coqdocvar{x} \coqdocvar{t1} \coqdocvar{t2}, \{\coqdocvar{x} := \coqdocvar{u}\}(\coqdocvar{n\_sub} \coqdocvar{t1} \coqdocvar{x} \coqdocvar{t2}) = \coqdocvar{n\_sub} \coqdocvar{t1} \coqdocvar{x} (\{\coqdocvar{x} := \coqdocvar{u}\}\coqdocvar{t2}).\coqdoceol
\coqdocemptyline
\coqdocemptyline
\end{coqdoccode}
We will now prove some stability results for the metasubstitution w.r.t. $\alpha$-equivalence. More precisely, we will prove that if $t =_\alpha t'$ and $u =_\alpha u'$ then $\metasub{t}{x}{u} =_\alpha \metasub{t'}{x}{u'}$, where $x$ is a variable and $t, t', u$ and $u'$ are terms. This proof is split in two cases: firstly, we prove that if $u =_\alpha u'$ then $\metasub{t}{x}{u} =_\alpha \metasub{t}{x}{u'}, \forall x, t, u, u'$; secondly, we prove that if $t =_\alpha t'$ then $\metasub{t}{x}{u} =_\alpha \metasub{t'}{x}{u}, \forall x, t, t', u$. These two cases are then combined through the transitivity of the $\alpha$-equivalence relation. Nevertheless, this task was not straighforward. Let's follow the steps of our first trial.
\begin{coqdoccode}
\coqdocemptyline
\coqdocnoindent
\coqdockw{Lemma} \coqdocvar{aeq\_m\_subst\_in\_trial}: \coqdockw{\ensuremath{\forall}} \coqdocvar{t} \coqdocvar{u} \coqdocvar{u'} \coqdocvar{x}, \coqdocvar{u} =\coqdocvar{a} \coqdocvar{u'} \ensuremath{\rightarrow} (\{\coqdocvar{x} := \coqdocvar{u}\}\coqdocvar{t}) =\coqdocvar{a} (\{\coqdocvar{x} := \coqdocvar{u'}\}\coqdocvar{t}).\coqdoceol
\end{coqdoccode}
\noindent{\bf Proof.} The proof is done by induction on the size of term \coqdocvar{t}, and we will focus on the abstraction case, {\it i.e.} $t = \lambda_y.t_1$. The goal in this case is $\metasub{(\lambda_y.t_1)}{x}{u} =_\alpha \metasub{(\lambda_y.t_1)}{x}{u'}$. \begin{enumerate} \item If $x = y$ then the result is trivial by lemma $m\_subst\_abs\_eq$. \item If $x \neq y$ then we need two fresh names in order to propagate the metasubstitution inside the abstractions on each side of the $\alpha$-equation. Let $x_0$ be a fresh name not in the set $fv\_nom(u) \cup fv\_nom(\lambda_y. t_1)\cup \{x\}$, and $x_1$ be a fresh name not in the set $fv\_nom(u') \cup fv\_nom(\lambda_y. t_1)\cup \{x\}$. After propagating the metasubstitution we need to prove $\lambda_{x_0}.\metasub{(\swap{y}{x_0}{t_1})}{x}{u} =_\alpha \lambda_{x_1}.\metasub{(\swap{y}{x_1}{t_1})}{x}{u'}$, and we proceed by comparing $x_0$ and $x_1$: \begin{enumerate} \item If $x_0 = x_1$ then we are done by the induction hypothesis. \item Otherwise, we need to apply the rule $aeq\_abs\_diff$ and the goal is $\metasub{(\swap{y}{x_0}{t_1})}{x}{u} =_\alpha \swap{x_0}{x_1}{(\metasub{(\swap{y}{x_1}{t_1})}{x}{u'})}$. But in order to proceed we need to know how to propagate the swap inside the metasubstitution, which is the content of the following lemma: \end{enumerate}\end{enumerate}
\begin{coqdoccode}
\coqdocnoindent
\coqdockw{Lemma} \coqdocvar{swap\_m\_subst}: \coqdockw{\ensuremath{\forall}} \coqdocvar{t} \coqdocvar{u} \coqdocvar{x} \coqdocvar{y} \coqdocvar{z}, \coqdocvar{swap} \coqdocvar{y} \coqdocvar{z} (\{\coqdocvar{x} := \coqdocvar{u}\}\coqdocvar{t}) =\coqdocvar{a} (\{(\coqdocvar{vswap} \coqdocvar{y} \coqdocvar{z} \coqdocvar{x}) := (\coqdocvar{swap} \coqdocvar{y} \coqdocvar{z} \coqdocvar{u})\}(\coqdocvar{swap} \coqdocvar{y} \coqdocvar{z} \coqdocvar{t})).\coqdoceol
\end{coqdoccode}
\noindent{\bf Proof.} We write the statement of the lemma in metanotation before starting the proof:\newline

$\forall t\ u\ x\ y\ z, \swap{y}{z}{(\metasub{t}{x}{u})} =_{\alpha} \metasub{\swap{y}{z}{t}}{\vswap{y}{z}{x}}{\swap{y}{z}{u}}$\newline

 The proof is by induction on the size of the term \coqdocvar{t}, and again we will focus only on the abstraction case, {\it i.e.} when $t = \lambda_w. t_1$. The goal in this case is $\swap{y}{z}{(\metasub{(\lambda_w.t_1)}{x}{u})} =_{\alpha} \metasub{(\swap{y}{z}{\lambda_w.t_1})}{\vswap{y}{z}{x}}{\swap{y}{z}{u}}$, and we proceed by comparing $x$ and $w$. \begin{enumerate} \item If $x = w$ the $\alpha$-equality is trivial. \item If $x \neq w$ then we need a fresh name, say $w_0$, to be able to propagate the metasubstitution inside the abstraction on the LHS of the $\alpha$-equation. The variable $w_0$ is taken such that it is not in the set $fv\_nom(u)\cup fv\_nom(\lambda_w. t_1) \cup \{x\}$, and we get the goal $\lambda_{\vswap{y}{z}{w_0}}.\swap{y}{z}{(\metasub{\swap{w}{w_0}{t_1}}{x}{u})} =_{\alpha} \metasub{(\lambda_{\vswap{y}{z}{w}}.\swap{y}{z}{t_1})}{\vswap{y}{z}{x}}{\swap{y}{z}{u}}$. Now we propagate the metasubstitution over the abstraction in the RHS of the goal. Since $x\neq w$ implies $\vswap{y}{z}{x} \neq \vswap{y}{z}{w}$, we need another fresh name, say $w_1$, not in the set $fv\_nom(\swap{y}{z}{u}) \cup fv\_nom(\lambda_{\vswap{y}{z}{w}}.\swap{y}{z}{t_1}) \cup \{\vswap{y}{z}{x}\}$, and after the propagation we need to prove that $\lambda_{\vswap{y}{z}{w_0}}.\swap{y}{z}{(\metasub{\swap{w}{w_0}{t_1}}{x}{u})} =_{\alpha} \lambda_{w_1}.\metasub{(\swap{w_1}{\vswap{y}{z}{w}}{(\swap{y}{z}{t_1})})}{\vswap{y}{z}{x}}{\swap{y}{z}{u}}$. We consider two cases: either $w_1 = \vswap{y}{z}{w_0}$ or $w_1 \neq \vswap{y}{z}{w_0}$. In the former case, we apply the rule $\mbox{\it aeq}\_\mbox{\it abs}\_\mbox{\it same}$ and we are done by the induction hypothesis. When $w_1 \neq \vswap{y}{z}{w_0}$, the application of the rule $\mbox{\it aeq}\_\mbox{\it abs}\_\mbox{\it diff}$ generates the goal

\begin{equation}\label{ext:swap}\swap{w_1}{\vswap{y}{z}{w_0}}{\swap{y}{z}{(\metasub{\swap{w}{w_0}{t_1}}{x}{u})}} =_{\alpha} \metasub{(\swap{w_1}{\vswap{y}{z}{w}}{(\swap{y}{z}{t_1})})}{\vswap{y}{z}{x}}{\swap{y}{z}{u}} \end{equation}

We can use the induction hypothesis to propagate the swap inside the metasubstitution, and then we get an $\alpha$-equality with metasubstitution as main operation on both sides, whose corresponding components are $\alpha$-equivalent. In a more abstract way, we have to prove an $\alpha$-equality of the form $\metasub{t}{x}{u} =_\alpha \metasub{t'}{x}{u'}$, where $t =_\alpha t'$ and $u =_\alpha u'$, but this is exactly what we were trying to prove  in the previous lemma. \end{enumerate} Therefore, we are in a circular problem because both \coqdocvar{aeq\_m\_subst\_in\_trial} and \coqdocvar{swap\_m\_subst} depend on each other to be proved!

Our solution to this problem consists in taking advantage of the fact that $\alpha$-equivalent terms have the same set of free variables (see lemma \coqdocvar{aeq\_fv\_nom}), and noting that the external swap in the LHS of (\ref{ext:swap}) was generated by the application of the rule \coqdocvar{aeq\_abs\_diff} because the abstractions have different bindings. Let's go back to the proof of lemma \coqdocvar{aeq\_m\_subst\_in}: 
\begin{coqdoccode}
\coqdocnoindent
\coqdockw{Lemma} \coqdocvar{aeq\_m\_subst\_in}: \coqdockw{\ensuremath{\forall}} \coqdocvar{t} \coqdocvar{u} \coqdocvar{u'} \coqdocvar{x}, \coqdocvar{u} =\coqdocvar{a} \coqdocvar{u'} \ensuremath{\rightarrow} (\{\coqdocvar{x} := \coqdocvar{u}\}\coqdocvar{t}) =\coqdocvar{a} (\{\coqdocvar{x} := \coqdocvar{u'}\}\coqdocvar{t}).\coqdoceol
\end{coqdoccode}
\noindent{\bf Proof.} We go directly to the abstraction case. When $t = \lambda_y.t_1$, the goal is $\metasub{(\lambda_y.t_1)}{x}{u} =_\alpha \metasub{(\lambda_y.t_1)}{x}{u'}$. If $x \neq y$ then the fresh name needed for the LHS must not belong to the set $fv\_nom(u) \cup fv\_nom(\lambda_y. t_1)\cup \{x\}$, while the fresh name for the RHS must not belong to $fv\_nom(u' ) \cup fv\_nom(\lambda_y. t_1)\cup \{x\}$. These sets differ only by the subsets $fv\_nom(u)$ and $fv\_nom(u' )$. Nevertheless, these subsets are equal because $u$ and $u'$ are $\alpha$-equivalent (see lemma \coqdocvar{aeq\_fv\_nom}). Concretely, the current goal is as follows:

\begin{verbatim}
 (let (z, _) := atom_fresh (union (fv_nom u) (union (fv_nom (n_abs y t1))
         (singleton x))) in n_abs z (subst_rec_fun (swap y z t1) u x)) =a
 (let (z, _) := atom_fresh (union (fv_nom u') (union (fv_nom (n_abs y t1))
            (singleton x))) in n_abs z (subst_rec_fun (swap y z t1) u' x))
\end{verbatim}

\noindent where the sets $fv\_nom(u)$ and $fv\_nom(u')$ appear in different $let$ expressions, each one is responsible for generating one fresh name. But since these sets are equal, if one could replace $fv\_nom(u)$ by $fv\_nom(u')$ (or vice-versa) then only one fresh name is generated after evaluating the \coqdocvar{atom\_fresh} function. Nevertheless, the only way that we managed to do such replacement was by adding the following axiom:

\begin{verbatim}
Axiom Eq_implies_equality: forall t1 t2, t1 =a t2 -> fv_nom t1 = fv_nom t2.
\end{verbatim}

This axiom is similar to lemma \coqdocvar{aeq\_fv\_nom} where the set equality [=] was replaced by the syntactic (Leibniz) equality =. Now, we can generate just one fresh name and propagate the metasubstitution on both sides of the goal, and we are done by the induction hypothesis. The case of the explicit substitution is similar, and with this strategy we avoid both the rules $\mbox{\it aeq}\_\mbox{\it abs}\_\mbox{\it diff}$ and $\mbox{\it aeq}\_\mbox{\it sub}\_\mbox{\it diff}$ that introduce swappings. $\hfill\Box$
\begin{coqdoccode}
\coqdocemptyline
\coqdocemptyline
\end{coqdoccode}
The next lemma, named \coqdocvar{aeq\_m\_subst\_out} will benefit the strategy used in the previous proof, but it is not straightfoward.
\begin{coqdoccode}
\coqdocemptyline
\coqdocnoindent
\coqdockw{Lemma} \coqdocvar{aeq\_m\_subst\_out}: \coqdockw{\ensuremath{\forall}} \coqdocvar{t} \coqdocvar{t'} \coqdocvar{u} \coqdocvar{x}, \coqdocvar{t} =\coqdocvar{a} \coqdocvar{t'} \ensuremath{\rightarrow} (\{\coqdocvar{x} := \coqdocvar{u}\}\coqdocvar{t}) =\coqdocvar{a} (\{\coqdocvar{x} := \coqdocvar{u}\}\coqdocvar{t'}).\coqdoceol
           \end{coqdoccode}
\noindent{\bf Proof.} The proof is by induction on the size of the term \coqdocvar{t}. Note that induction on the hypothesis \coqdocvar{t} =\coqdocvar{a} \coqdocvar{t'} does not work due to a similar problem involving swaps that appears when structural induction on \coqdocvar{t} is used. The abstraction and the explicit substitution are the interesting cases.

In the abstraction case, we need to prove that $\metasub{(\lambda_y.t_1)}{x}{u} =_{\alpha} \metasub{t'}{x}{u}$, where $\lambda_y. t_1 =_{\alpha} t'$ by hypothesis. Therefore, $t'$ must be an abstraction, and according to our definition of $\alpha$-equivalence there are two possible subcases: \begin{enumerate} \item In the first subcase, $t' = \lambda_y.t_2$, where $t_1 =_{\alpha} t_2$, and hence the current goal is $\metasub{(\lambda_y.t_1)}{x}{u} =_{\alpha} \metasub{(\lambda_y.t_2)}{x}{u}$. We proceed by comparing $x$ and $y$:
\begin{enumerate}
 \item If $x = y$ then, we are done by using twice lemma $m\_subst\_abs\_eq$.
 \item When $x \neq y$, then we need to propagate the metasubstitution on both sides of the goal. On the LHS, we need a fresh name that is not in the set $fv(u)\cup fv(\lambda_y.t_1) \cup \{x\}$, while for the RHS, the fresh name cannot belong to the set $fv(u)\cup fv(\lambda_y.t_2) \cup \{x\}$. From the hypothesis $t_1 =_{\alpha} t_2$, we know, by lemma $aeq\_fv\_nom$, that the sets $fv\_nom(t_1)$ and  $fv\_nom(t_2)$ are equal. Therefore, we can take just one fresh name, say $z$, and propagate both metasubstitutions over abstractions with the same binding, and we conclude with the induction hypothesis.
\end{enumerate}
\item In the second subcase, $t' = \lambda_{y_0}.t_2$, where $t_1 =_{\alpha} \swap{y_0}{y}{t_2}$ and $y \neq y_0$. The current goal is $$\metasub{(\lambda_y.t_1)}{x}{u} =_{\alpha} \metasub{(\lambda_{y_0}.t_2)}{x}{u}$$ and we proceed by comparing $x$ and $y$:
\begin{enumerate}
 \item If $x = y$ then the goal simplifies to $\lambda_y.t_1 =_{\alpha} \metasub{(\lambda_{y_0}.t_2)}{x}{u}$ by lemma $m\_subst\_abs\_eq$, and we pick a fresh name $x$, that is not in the set $fv\_nom(u) \cup fv\_nom(\lambda_{y_0}.t_2) \cup \{y\}$, and propagate the metasubstitution on the RHS of the goal, resulting in the new goal $\lambda_y. t_1 =_{\alpha} \lambda_x.\metasub{(\swap{y_0}{x}{t_2})}{y}{u}$. Note that the metasubstitution on the RHS has no effect in the term $\swap{y_0}{x}{t_2}$ because $y \neq y_0$, $y \neq x$ and $y$ does not occur free in $t_2$ and we conclude by hypothesis.
\item If $x \neq y$ then we proceed by comparing $x$ and $y_0$ on the RHS, and the proof, when $x = y_0$, is analogous to the previous subcase. When both $x \neq y$ and $x \neq y_0$ then we need to propagate the metasubstitution on both sides of the goal $\metasub{(\lambda_y.t_1)}{x}{u} =_{\alpha} \metasub{(\lambda_{y_0}.t_2)}{x}{u}$. We have that $\lambda_y.t_1 =_{\alpha} \lambda_{y_0}.t_2$ and hence the sets $fv\_nom(\lambda_y.t_1)$ and $fv\_nom(\lambda_{y_0}.t_2)$ are equal. Therefore, only one fresh name, say $x_0$, that is not in the set $x_0 \notin fv\_nom(u) \cup fv\_nom(\lambda_{y_0}.t_2) \cup \{x\}$ is enough to fulfill the conditions for propagating the metasubstitutions on both sides of the goal, and we are done by the induction hypothesis.
\end{enumerate}
\item The explicit substitution operation is also interesting, but we will not comment because we are running out of space. $\hfill\Box$
\end{enumerate}

 As a corollary, one can join the lemmas \coqdocvar{aeq\_m\_subst\_in} and \coqdocvar{aeq\_m\_subst\_out} as follows:
\begin{coqdoccode}
\coqdocemptyline
\coqdocnoindent
\coqdockw{Corollary} \coqdocvar{aeq\_m\_subst\_eq}: \coqdockw{\ensuremath{\forall}} \coqdocvar{t} \coqdocvar{t'} \coqdocvar{u} \coqdocvar{u'} \coqdocvar{x}, \coqdocvar{t} =\coqdocvar{a} \coqdocvar{t'} \ensuremath{\rightarrow} \coqdocvar{u} =\coqdocvar{a} \coqdocvar{u'} \ensuremath{\rightarrow} (\{\coqdocvar{x} := \coqdocvar{u}\}\coqdocvar{t}) =\coqdocvar{a} (\{\coqdocvar{x} := \coqdocvar{u'}\}\coqdocvar{t'}).\coqdoceol
\coqdocemptyline
\end{coqdoccode}
Now, we show how to propagate a swap inside metasubstitutions using the decomposition of the metasubstitution provided by the corollary \coqdocvar{aeq\_m\_subst\_eq}.\newline 
\begin{coqdoccode}
\coqdocemptyline
\coqdocnoindent
\coqdockw{Lemma} \coqdocvar{swap\_subst\_rec\_fun}: \coqdockw{\ensuremath{\forall}} \coqdocvar{x} \coqdocvar{y} \coqdocvar{z} \coqdocvar{t} \coqdocvar{u}, \coqdocvar{swap} \coqdocvar{x} \coqdocvar{y} (\{\coqdocvar{z} := \coqdocvar{u}\}\coqdocvar{t}) =\coqdocvar{a} (\{(\coqdocvar{vswap} \coqdocvar{x} \coqdocvar{y} \coqdocvar{z}) := (\coqdocvar{swap} \coqdocvar{x} \coqdocvar{y} \coqdocvar{u})\}(\coqdocvar{swap} \coqdocvar{x} \coqdocvar{y} \coqdocvar{t})).\coqdoceol
\end{coqdoccode}
\noindent{\bf Proof.} Firstly, we write the lemma in metanotation: $\forall x\ y\ z\ t\ u, \swap{x}{y}{\metasub{t}{z}{u}} =_{\alpha} \metasub{\swap{x}{y}{t}}{\vswap{x}{y}{z}}{\swap{x}{y}{u}}$. Next, we compare $x$ and $y$, since the case $x = y$ is trivial. When $x \neq y$, the proof proceeds by induction on the size of the term $t$. The tricky cases are the abstraction and explicit substitution, and we comment just the former case. If $t = \lambda_{y'}.t_1$ then we must prove that $\swap{x}{y}{\metasub{(\lambda_{y'}.t_1)}{z}{u} =_{\alpha} \metasub{\swap{x}{y}{(\lambda{y'}.t_1)}}{\vswap{x}{y}{z}}{\swap{x}{y}{u}}}$. Firstly, we compare the variables $y'$ and $z$ according to the definition of the metasubstitution:                                          \begin{enumerate}
\item When $y' = z$ the metasubstitution is erased according to the definition (\ref{msubst}) on both sides of the goal and we are done.
\item When $y' \neq z$ then the metasubstitutions on both sides of the goal need to be propagated inside the corresponding abstractions. In order to do so, a new name need to be created. Note that in this case, it is not possible to create a unique name for both sides because the two sets are different. In fact, in the LHS the fresh name cannot belong to the set $fv\_nom(\lambda_y'.t_1) \cup fv\_nom(u) \cup \{z\}$, while the name of the RHS cannot belong to the set $fv\_nom(\swap{x}{y}{\lambda_y'.t_1}) \cup fv\_nom(\swap{x}{y}{u}) \cup \{\vswap{x}{y}{z}\}$. Let $x_0$ be a fresh name that is not in the set $fv\_nom(\lambda_y'.t_1) \cup fv\_nom(u) \cup \{z\}$, and $x_1$ a fresh name that is not in the set $fv\_nom(\swap{x}{y}{\lambda_y'.t_1}) \cup fv\_nom(\swap{x}{y}{u}) \cup \{\vswap{x}{y}{z}\}$. After the propagation of the metasubstitutions, we have to prove that $\lambda_{\vswap{x}{y}{x0}}.(\swap{x}{y}{(\metasub{(\swap{y'}{x_0}{t_1})}{z}{u})} =_{\alpha} \lambda_{x_1}.(\metasub{\swap{(\vswap{x}{y}{y'})}{x_1}{(\swap{x}{y}{t_1})}}{\vswap{x}{y}{z}}{\swap{x}{y}{u}})$. We proceed by comparing $x_1$ with $\vswap{x}{y}{x_0}$.
\begin{enumerate}
\item If $x_1 = \vswap{x}{y}{x_0}$ then we use the induction hypothesis to propagate the swap inside the metasubstitution in the LHS, and we get the goal $\metasub{(\swap{x}{y}{(\swap{y'}{x_0}{t_1})})}{\vswap{x}{y}{z}}{\swap{x}{y}{u}}  =_{\alpha}
  \metasub{(\swap{(\vswap{x}{y}{y'})}{(\vswap{x}{y}{x_0})}{(\swap{x}{y}{t_1})})}{\vswap{x}{y}{z}}{\swap{x}{y}{u}}$ that is proved by the swap equivariance lemma $swap\_equivariance$.
 \item If $x_1 \neq \vswap{x}{y}{x_0}$ then by the rule $aeq\_abs\_diff$ we have to prove that the variable $\vswap{x}{y}{x_0}$ is not in the set of free variables of the term $\metasub{(\swap{\vswap{x}{y}{y'}}{x_1}{\swap{x}{y}{t_1}})}{\vswap{x}{y}{z}}{\swap{x}{y}{u}}$ and that $\swap{x}{y}{(\metasub{(\swap{y'}{x_0}{t_1})}{z}{u})} =_{\alpha}
  \swap{x_1}{(\vswap{x}{y}{x_0})}{(\metasub{(\swap{(\vswap{x}{y}{y'})}{x_1}{(\swap{x}{y}{t_1})})}{\vswap{x}{y}{z}}{\swap{x}{y}{u}}}$. The former condition is routine. The later condition is proved using the induction hypothesis twice to propagate the swaps inside the metasubstitutions on each side of the $\alpha$-equality. This swap has no effect on the variable $z$ of the metasubstitution because $x_1$ is different from $\vswap{x}{y}{z}$, and $x_0$ is different from $z$. Therefore we can apply lemma $aeq\_m\_subst\_eq$, and each generated case is proved by routine manipulation of swaps.
\end{enumerate} $\hfill\Box$
\end{enumerate}

 The following two lemmas toghether with lemmas \coqdocvar{m\_subst\_abs\_eq} and \coqdocvar{m\_subst\_sub\_eq} are essential in simplifying the propagations of metasubstitution. They are presented here because they depend on lemma \coqdocvar{swap\_subst\_rec\_fun}. 
\begin{coqdoccode}
\coqdocemptyline
\coqdocnoindent
\coqdockw{Lemma} \coqdocvar{m\_subst\_abs\_neq}: \coqdockw{\ensuremath{\forall}} \coqdocvar{t} \coqdocvar{u} \coqdocvar{x} \coqdocvar{y} \coqdocvar{z}, \coqdocvar{x} \ensuremath{\not=} \coqdocvar{y} \ensuremath{\rightarrow} \coqdocvar{z} `\coqdocvar{notin}` \coqdocvar{fv\_nom} \coqdocvar{u} `\coqdocvar{union}` \coqdocvar{fv\_nom} (\coqdocvar{n\_abs} \coqdocvar{y} \coqdocvar{t}) `\coqdocvar{union}` \{\{\coqdocvar{x}\}\} \ensuremath{\rightarrow} \{\coqdocvar{x} := \coqdocvar{u}\}(\coqdocvar{n\_abs} \coqdocvar{y} \coqdocvar{t}) =\coqdocvar{a} \coqdocvar{n\_abs} \coqdocvar{z} (\{\coqdocvar{x} := \coqdocvar{u}\}(\coqdocvar{swap} \coqdocvar{y} \coqdocvar{z} \coqdocvar{t})).\coqdoceol
\coqdocemptyline
\coqdocnoindent
\coqdockw{Lemma} \coqdocvar{m\_subst\_sub\_neq} : \coqdockw{\ensuremath{\forall}} \coqdocvar{t1} \coqdocvar{t2} \coqdocvar{u} \coqdocvar{x} \coqdocvar{y} \coqdocvar{z}, \coqdocvar{x} \ensuremath{\not=} \coqdocvar{y} \ensuremath{\rightarrow} \coqdocvar{z} `\coqdocvar{notin}` \coqdocvar{fv\_nom} \coqdocvar{u} `\coqdocvar{union}` \coqdocvar{fv\_nom} ([\coqdocvar{y} := \coqdocvar{t2}]\coqdocvar{t1}) `\coqdocvar{union}` \{\{\coqdocvar{x}\}\} \ensuremath{\rightarrow} \{\coqdocvar{x} := \coqdocvar{u}\}([\coqdocvar{y} := \coqdocvar{t2}]\coqdocvar{t1}) =\coqdocvar{a} ([\coqdocvar{z} := (\{\coqdocvar{x} := \coqdocvar{u}\}\coqdocvar{t2})](\{\coqdocvar{x} := \coqdocvar{u}\}(\coqdocvar{swap} \coqdocvar{y} \coqdocvar{z} \coqdocvar{t1}))).\coqdoceol
\coqdocemptyline
\coqdocemptyline
\coqdocemptyline
\coqdocemptyline
\end{coqdoccode}
In the pure $\lambda$-calculus, the substitution lemma is probably the first non trivial property. In our framework, we have defined two different substitution operators, namely, the metasubstitution denoted by $\metasub{t}{x}{u}$ and the explicit substitution, written as $\esub{t}{x}{u}$. In what follows, we present the main steps of our proof of the substitution lemma for \coqdocvar{n\_sexp} terms, {\it i.e.} for nominal terms with explicit substitutions. 
\begin{coqdoccode}
\coqdocemptyline
\coqdocnoindent
\coqdockw{Lemma} \coqdocvar{m\_subst\_lemma}: \coqdockw{\ensuremath{\forall}} \coqdocvar{t1} \coqdocvar{t2} \coqdocvar{t3} \coqdocvar{x} \coqdocvar{y}, \coqdocvar{x} \ensuremath{\not=} \coqdocvar{y} \ensuremath{\rightarrow} \coqdocvar{x} `\coqdocvar{notin}` (\coqdocvar{fv\_nom} \coqdocvar{t3}) \ensuremath{\rightarrow}\coqdoceol
\coqdocindent{10.50em}
(\{\coqdocvar{y} := \coqdocvar{t3}\}(\{\coqdocvar{x} := \coqdocvar{t2}\}\coqdocvar{t1})) =\coqdocvar{a} (\{\coqdocvar{x} := (\{\coqdocvar{y} := \coqdocvar{t3}\}\coqdocvar{t2})\}(\{\coqdocvar{y} := \coqdocvar{t3}\}\coqdocvar{t1})).\coqdoceol
\end{coqdoccode}
\noindent{\bf Proof.} The proof is by induction on the size of \coqdocvar{t1}. The interesting cases are the abstraction and the explicit substitution. We focus on the former, {\it i.e.} $t1= \lambda_z.t_1'$, whose initial goal is

$\metasub{(\metasub{(\lambda_z.t_1')}{x}{t_2})}{y}{t_3} =_{\alpha} \metasub{(\metasub{(\lambda_z .t_1')}{y}{t_3})}{x}{\metasub{t_2}{y}{t_3}}$

\noindent assuming that $x \neq y$ and $x \notin fv\_nom(t_3)$. The induction hypothesis generated by this case states that the lemma holds for any term of the size of $t_1'$, {\it i.e.} any term with the same size of the body of the abstraction. We start comparing $z$ with $x$ aiming to apply the definition of the metasubstitution on the LHS of the goal. \begin{enumerate}
\item When $z = x$, the subterm $\metasub{\lambda_x.t_1'}{x}{t_2}$ reduces to $\lambda_x.t_1'$ by lemma $m\_subst\_abs\_eq$, and then the LHS reduces to $\metasub{\lambda_x.t_1'}{y}{t_3}$. The RHS $\metasub{\metasub{\lambda_x.t_1'}{y}{t_3}}{x}{\metasub{t_2}{y}{t_3}}$ also reduces to it because $x$ does not occur free neither in $\lambda_x.t_1'$ nor in $t_3$, and we are done.
\item When $z \neq x$, then we compare $y$ with $z$.
\begin{enumerate}
 \item When $y = z$, the subterm $\metasub{(\lambda_z.t_1')}{y}{t_3}$ can be simplified to $\lambda_z.t_1'$, by lemma $m\_subst\_abs\_eq$. On the LHS, we propagate the internal metasubstitution over the abstraction taking a fresh name $w$ not in the set $fv\_nom(\lambda_z.t_1') \cup fv\_nom(t_3) \cup fv\_nom(t_2) \cup \{x\}$, where the goal is $\metasub{(\lambda_w.(\metasub{\swap{z}{w}{t_1'}}{x}{t_2}))}{z}{t_3} =_{\alpha} \metasub{(\lambda_z.t_1')}{x}{\metasub{t_2}{z}{t_3}}$. We proceed by comparing $z$ and $w$: \begin{enumerate}
\item If $z = w$ then the current goal simplifies to

$\metasub{(\lambda_w.(\metasub{t_1'}{x}{t_2}))}{w}{t_3} =_{\alpha} \metasub{(\lambda_w.t_1')}{x}{\metasub{t_2}{w}{t_3}}$

We can propagate the metasubstitution on the RHS and there is no need for a fresh name since the variable $w$ fullfil the condition required by lemma $m\_subst\_abs\_neq$. We conclude with lemmas $aeq\_m\_subst\_in$ and $m\_subst\_notin$.
\item If $z \neq w$ then we can propagate the metasubstitutions on both sides of the goal taking $w$ as the fresh name that fullfil the conditions of lemma $m\_subst\_abs\_neq$. We proceed with $aeq\_abs\_same$, and conclude by the induction hypothesis.
\end{enumerate}
\item If $y \neq z$ then we follow a similar strategy that avoids unnecessary generation of fresh names. In this way, we take a fresh $w$ that is not in the set $fv\_nom(t_3)\cup fv\_nom(t_2)\cup fv\_nom(\lambda_z.t_1')\cup \{x\}\cup \{y\}$, and propagate the metasubstitution inside the abstraction resulting in the goal $\lambda_w. (\metasub{(\metasub{\swap{z}{w}{t_1'}}{x}{t_2})}{y}{t_3} =_{\alpha} \lambda_w. (\metasub{(\metasub{\swap{z}{w}{t_1'}}{y}{t_3})}{x}{\metasub{t_2}{y}{t_3}}$. We conclude by the induction hypothesis. $\hfill\Box$
\end{enumerate}
\end{enumerate}
\newpage

\section{Conclusion and Future work}

In this work, we presented a formalization of the substitution lemma in a framework that extends the $\lambda$-calculus with an explicit substitution operator. Calculi with explicit substitutions are important frameworks to study properties of the $\lambda$-calculus and have been extensively studied in the last decades \cite{abadiExplicitSubstitutions1991,accattoliAbstractFactorizationTheorem2012a,ayala-rinconComparingCalculiExplicit2002a,ayala-rinconComparingImplementingCalculi2005a,bonelliPerpetualityNamedLambda2001}. 
 
The formalization is modular in the sense that the explicit substitution operator is generic and could be instantiated with any calculi with  explicit substitutions in a nominal setting. Despite the fact that our definition of metasubstitution, called $subst\_rec\_fun$, performs a renaming with a fresh name whenever it is propagated inside a binding structure (either an abstraction or an explicit substitution in our case), we showed how to avoid unnecessary generation of fresh names that could result in a circular problem in the proofs. Several auxiliary (minor) results were not included in this document, but they are numerous and can be found in the source file of the formalization that is publicly available at \url{https://flaviomoura.info/files/msubst.v}

As future work, we intend to get rid of the axiom $Eq\_implies\_equality$. The natural candidate for this would be the use of generalized rewriting, {\it i.e.} setoid rewriting, but it not clear whether generalized rewriting allows a rewrite step in a let expression. Another possibility is the implementation of the metasubstitution using recursors \cite{popescuNominalRecursorsEpiRecursors2023,gheriFormalizedGeneralTheory2020}. In addition, we plan to integrate this formalization with another one related to the Z property\footnote{\url{https://cicm-conference.org/2021/cicm.php?event=fmm&menu=general}} to prove confluence  of calculi with explicit substitutions \cite{nakazawaCompositionalConfluenceProofs2016,nakazawaPropertyShufflingCalculus2023}, as well as  other properties in the nominal framework \cite{kesnerPerpetualityFullSafe2008}.

\nocite{*}
\bibliographystyle{eptcs}
\bibliography{from2023}
\end{document}